\documentclass[aps,twocolumn]{revtex4}
\usepackage{amsmath,amssymb}

\begin{document}

\title{Power-Law Entropic Corrections to Newton's Law and Friedmann Equations}
\author{A. Sheykhi $^{1,2}$\thanks{%
email address: hendi@mail.yu.ac.ir} and  S. H. Hendi $^{2,3} $\thanks{%
email address: sheykhi@mail.uk.ac.ir}} \affiliation{$^1$
Department of Physics, Shahid Bahonar University, P.O. Box 76175,
Kerman, Iran
\\
$^2$ Research Institute for Astronomy and Astrophysics of Maragha
(RIAAM),
Maragha, Iran\\
$^3$ Physics Department, College of Sciences, Yasouj University,
Yasouj 75914, Iran}

\begin{abstract}
A possible source of black hole entropy could be the entanglement of
quantum fields in and out the horizon. The entanglement entropy of
the ground state obeys the area law. However, a correction term
proportional to a fractional power of area results when the field is
in a superposition of ground and excited states. Inspired by the
power-law corrections to entropy and adopting the viewpoint that
gravity emerges as an entropic force, we derive modified Newton's
law of gravitation as well as the corrections to Friedmann
equations. In a different approach, we obtained power-law corrected
Friedmann equation by starting from the first law of thermodynamics
at apparent horizon of a FRW universe, and assuming that the
associated entropy with apparent horizon has a power-law corrected
relation. Our study shows a consistency between the obtained results
of these two approaches. We also examine the time evolution of the
total entropy including the power-law corrected entropy associated
with the apparent horizon together with the matter field entropy
inside the apparent horizon and show that the generalized second law
of thermodynamics is fulfilled in a region enclosed by the apparent
horizon.

\end{abstract}

 \maketitle
\section{Introduction}
Recently, Verlinde \cite{Verlinde} demonstrated that gravity can be
interpreted as an entropic force caused by the changes in the
information associated with the positions of material bodies. In his
new proposal, Verlinde obtained successfully the Newton's law of
gravitation, the Poisson's equation and Einstein field equations by
employing the holographic principle together with the equipartition
law of energy. As soon as Verlinde presented his idea, many relevant
works about entropic force appeared. For example, Friedmann
equations from entropic force have been derived in Refs.
\cite{Shu,Cai1}. The Newtonian gravity \cite{Smolin}, the
holographic dark energy \cite{Li} and thermodynamics of black holes
\cite{Tian} have been investigated by using the entropic force
approach. It has been shown that uncertainty principle may arise in
the entropic force paradigm \cite{Vancea}. Other studies on the
entropic force, which raised a lot of attention recently, have been
carried out in \cite{other}.

On the other hand, string theory, as well as the string inspired
braneworld scenarios such as RSII model, suggest a modification of
Newton's law of gravitation at small distance scales
\cite{Polchinski,Randall}. In addition, there have been considerable
works on quantum corrections to some basic physical laws. The loop
quantum corrections to the Newton and Coulomb potential have been
investigated in some references (see \cite{Donoghue} and references
therein). Also, corrections to Friedmann equations from loop quantum
gravity has been studied in \cite{Taveras}.

Inspired by  Verlinde's argument and considering the quantum
corrections to the area law of the black hole entropy, one is able
to derive some physical equations with correction terms. For
example, modified Newton's law of gravitation has been studied in
\cite{Modesto}, while, modified Friedmann equations have been
constructed in \cite{Sheykhi1, BLi}. In all these cases
\cite{Modesto,Sheykhi1, BLi} the corrected entropy has the
logarithmic term which arises from the inclusion of quantum effects,
motivated from the loop quantum gravity and is due to the thermal
equilibrium fluctuations and quantum fluctuations \cite{Rovelli}. In
addition, entropic corrections to Coulomb's law have also been
investigated in \cite{modifiedNC}. Very recently, by considering the
quantum corrections to the area law of black hole entropy,  the
modified forms of Poisson's equation of gravity, MOND theory of
gravitation and Einstein field equations were derived using the
entropic force interpretation of gravity \cite{hendisheykhi}.

In this paper we would like to consider the effects of the power-law
correction terms to the entropy on the Newton's law and Friedmann
equation. The power-law corrections to entropy appear in dealing
with the entanglement of quantum fields in and out the horizon
\cite{Sau}. Indeed, it has been shown that the origin of black hole
entropy may lie in the entanglement of quantum fields between inside
and outside of the horizon \cite{Sau}. Since the modes of
gravitational fluctuations in a black hole background behave as
scalar fields, one is able to compute the entanglement entropy of
such a field, by tracing over its degrees of freedom inside a
sphere. In this way the authors of \cite{Sau} showed that the black
hole entropy is proportional to the area of the sphere when the
field is in its ground state, but a correction term proportional to
a fractional power of area results when the field is in a
superposition of ground and excited states. For large horizon areas,
these corrections are relatively small and the area law is
recovered. Applying this power-law corrected entropy, we obtain the
corrections to  Newton's law as well as modified Friedmann equation
by adopting the viewpoint that gravity emerges as an entropic force.

The outline of our paper is as follows. In the next section, we use
Verlinde approach to derive Newton's law of gravitation with a
correction term resulting from the entanglement of quantum fields in
and out the horizon. In section \ref{Friedmann1}, we derive the
power-law entropy-corrected Friedmann equation of FRW universe by
considering gravity as an entropic force. Then, in section
\ref{Friedmann2}, we obtain modified Friedmann equation by applying
the first law of thermodynamics at apparent horizon of a FRW
universe. In section \ref{GSL} we examine to see whether the
power-law entropy-area relation together with the matter field
entropy inside the apparent horizon will satisfy the generalized
second law of thermodynamics. The last section is devoted to
conclusions and discussions.

\section{Entropic correction to Newton's law \label{Newton}}
According to Verlinde's argument, when a test particle moves apart
from the holographic screen, the magnitude of the entropic force
on this body has the form
\begin{equation}\label{F}
F\triangle x=T \triangle S,
\end{equation}
where $\triangle x$ is the displacement of the particle from the
holographic screen, while $T$ and $\triangle S$ are the
temperature and the entropy change on the screen, respectively.

In Verlinde's discussion, the black hole entropy $S$ plays a
significant role. Indeed, the derivation of Newton's law of
gravity depends on the entropy-area relationship $S=k_{B}A/4\ell
_p^2$ of black holes in Einstein's gravity, where $A =4\pi R^2$
represents the area of the horizon and $\ell _p=\sqrt{G\hbar/c^3}$
is the Planck length. However, the area law of black hole entropy
can be modified \cite{Sau}. The corrected entropy takes the form
\cite{pavon1}
\begin{equation}\label{S}
S=\frac{k_{B}A }{4\ell
_p^2}\left[1-K_{\alpha}A^{1-\alpha/2}\right],
\end{equation}
where $\alpha$ is a dimensionless constant whose value is
currently under debate, $k_{B}$ stands for the Boltzmann constant
and
\begin{equation}
\label{K} K_{\alpha}=\frac{\alpha
(4\pi)^{\alpha/2-1}}{(4-\alpha)r_c^{2-\alpha}},
\end{equation}
where $r_c$ is the crossover scale. The second term in the above
Eq. (\ref{S}) may be regarded as a power law correction to the
area law, resulting from entanglement, when the wave-function of
the field is chosen to be a superposition of ground state and
exited states.

Considering  the power-law correction to entropy, we show that
Newton's law of gravitation as well as Friedman equations will be
modified accordingly. First of all, we rewrite Eq. (\ref{S}) in
the following form
\begin{equation}
\label{S2}
 S=k_{B}\left[\frac{ A}{4\ell _p^2}+{s}(A)\right],
\end{equation}
where $s(A)$ stands for the correction term in the entropy
expression. Suppose we have two masses one a test mass and the
other considered as the source with respective masses $m$ and $M$.
Centered around the source mass $M$, is a spherically symmetric
surface $\mathcal {S}$ which will be defined with certain
properties that will be made explicit later. To derive the
entropic law, the surface $\mathcal {S}$ is between the test mass
and the source mass, but the test mass is assumed to be very close
to the surface as compared to its reduced Compton wavelength
$\lambda_m=\frac{\hbar}{mc}$. When a test mass $m$ is a distance
$\triangle x = \eta \lambda_m$ away from the surface $\mathcal
{S}$, the entropy of the surface changes by one fundamental unit
$\triangle S$ fixed by the discrete spectrum of the area of the
surface via the relation
\begin{equation}
\label{S3}
 \triangle S=\frac{\partial S}{\partial A}\triangle A=k_{B}\left(\frac{1}{4\ell _p^2}+\frac{\partial{s}(A)}{\partial A}\right)\triangle
 A.
\end{equation}
The energy of the surface $\mathcal {S}$ is identified with the
relativistic rest mass of the source mass:
\begin{equation}
\label{Ec} E=M c^2.
\end{equation}
On the surface $\mathcal {S}$, there live a set of ``bytes" of
information that scale proportional to the area of the surface so
that
\begin{equation}
\label{AQN}
 A=QN,
 \end{equation}
where $N$ represents the number of bytes and $Q$ is a fundamental
constant which should be specified later. Assuming the temperature
on the surface is $T$, and then according to the equipartition law
of energy \cite{Pad1}, the total energy on the surface is
\begin{equation}
\label{E}
 E=\frac{1}{2}Nk_B T.
 \end{equation}
Finally, we assume that the force on the particle follows from the
generic form of the entropic force governed by the thermodynamic
equation
\begin{equation}\label{F2}
F=T \frac{\triangle S}{\triangle x},
\end{equation}
where $\triangle S$ is one fundamental unit of entropy when
$|\triangle x|= \eta \lambda_m$, and the entropy gradient points
radially from the outside of the surface to inside. Note that $N$ is
the number of bytes and thus we set $\triangle N=1$; hence from
(\ref{AQN}) we have $\triangle A=Q$. Combining Eqs. (\ref{S3})-
(\ref{F2}), we find
\begin{equation}\label{F3}
F=-\frac{Mm}{R^2}\left(\frac{Q^2c^3}{8\pi \hbar \eta \ell
_p^2}\right)\left[1+4\ell_p^2\frac{\partial{s(A)}}{\partial
A}\right]_{A=4\pi R^2}.
\end{equation}
This is nothing but the Newton's law of gravitation to the first
order provided we define $Q^2=8\pi \eta \ell_p^4$. Thus we reach
\begin{equation}\label{F4}
F=-\frac{GMm}{R^2}\left[1+4\ell_p^2 \frac{\partial{s}}{\partial
A}\right]_{A=4\pi R^2}.
\end{equation}
Using Eq. (\ref{S}) we obtain
\begin{equation}\label{sder}
\left(\frac{\partial{s}}{\partial A}\right)_{A=4\pi R^2}=-\frac{
K_{\alpha}(4-\alpha)}{8\ell_p^2}\left(4\pi R^2\right)^{1-\alpha/2}
\end{equation}
Substituting Eq. (\ref{sder}) in Eq. (\ref{F4}) we obtain
\begin{equation}\label{F5}
F=-\frac{GMm}{R^2}\left[1-\frac{
K_{\alpha}}{2}(4-\alpha)\left(4\pi R^2\right)^{1-\alpha/2}\right],
\end{equation}
Using Eq. (\ref{K}) the above relation can be rewritten as
\begin{equation}\label{F5}
F=-\frac{GMm}{R^2}\left[1-\frac{\alpha}{2}\left(\frac{r_c}{R}\right)^{\alpha-2}\right],
\end{equation}
This is the power-law correction to the Newton's law of gravitation.
When $\alpha=0$, one recovers the usual Newton's law. Since gravity
is an attractive force we should have $F<0$. This requires
\begin{equation}\label{cond0}
1-\frac{\alpha}{2}\left(\frac{r_c}{R}\right)^{\alpha-2}>0,
\end{equation}
which can also be rewritten as
\begin{equation}\label{cond0}
\alpha<2 \left(\frac{R}{r_c}\right)^{\alpha-2},
\end{equation}
As we will see in section \ref{GSL}, this condition is also
necessary for satisfaction of the generalized second law of
thermodynamics for the universe with the power-law corrected
entropy.
\section{Entropic Corrections to Friedmann Equations \label{Friedmann1}}
Next, we extend our discussion to the cosmological setup. Assuming
the background spacetime to be spatially homogeneous and isotropic
which is described by the line element
\begin{equation}
ds^2={h}_{\mu \nu}dx^{\mu} dx^{\nu}+R^2(d\theta^2+\sin^2\theta
d\phi^2),
\end{equation}
where $R=a(t)r$, $x^0=t, x^1=r$, the two dimensional metric $h_{\mu
\nu}$=diag $(-1, a^2/(1-kr^2))$. Here $k$ denotes the curvature of
space with $k = 0, 1, -1$ corresponding to open, flat, and closed
universes, respectively. The dynamical apparent horizon, a
marginally trapped surface with vanishing expansion, is determined
by the relation $h^{\mu \nu}\partial_{\mu}R\partial_{\nu}R=0$. A
simple calculation gives the apparent horizon radius for the
Friedmann-Robertson-Walker (FRW) universe
\begin{equation}
\label{radius}
 R=ar=\frac{1}{\sqrt{H^2+k/a^2}},
\end{equation}
where $H=\dot{a}/a$ is the Hubble parameter. We also assume the
matter source in the FRW universe is a perfect fluid of mass
density $\rho$ and pressure $p$ with stress-energy tensor
\begin{equation}\label{T}
T_{\mu\nu}=(\rho+p)u_{\mu}u_{\nu}+pg_{\mu\nu}.
\end{equation}
Due to the pressure, the total mass $M = \rho V$ in the region
enclosed by the boundary $\mathcal S$ is no longer conserved, the
change in the total mass is equal to the work made by the pressure
$dM = -pdV$ , which leads to the well-known continuity equation
\begin{equation}\label{Cont}
\dot{\rho}+3H(\rho+p)=0,
\end{equation}
It is instructive to first derive the dynamical equation for
Newtonian cosmology. Consider a compact spatial region $V$ with a
compact boundary $\mathcal S$, which is a sphere with physical
radius $R= a(t)r$. Note that here $r$ is a dimensionless quantity
which remains constant for any cosmological object partaking in
free cosmic expansion. Combining the second law of Newton for the
test particle $m$ near the surface with gravitational force
(\ref{F5}) we get
\begin{equation}\label{F6}
F=m\ddot{R}=m\ddot{a}r=-\frac{GMm}{R^2}\left[1-\frac{\alpha}{2}\left(\frac{r_c}{R}\right)^{\alpha-2}\right],
\end{equation}
We also assume $\rho=M/V$ is the energy density of the matter
inside the the volume $V=\frac{4}{3} \pi a^3 r^3$. Thus, Eq.
(\ref{F6}) can be rewritten as
\begin{equation}\label{F7}
\frac{\ddot{a}}{a}=-\frac{4\pi
G}{3}\rho\left[1-\frac{\alpha}{2}\left(\frac{r_c}{R}\right)^{\alpha-2}\right],
\end{equation}
This is nothing but the power-law entropy-corrected dynamical
equation for Newtonian cosmology. The main difference between this
equation and the standard dynamical equation for Newtonian
cosmology is that the correction terms now depends explicitly on
the radius $R$. However, we can remove this confusion. Assuming
that for Newtonian cosmology the spacetime is Minkowskian with
$k=0$, then we get $R=1/H$, and we can rewrite Eq. (\ref{F7}) in
the form
\begin{equation}\label{F8}
\frac{\ddot{a}}{a}=-\frac{4\pi G}{3}\rho\left[1-\frac{\alpha
}{2}{r_c}^{\alpha-2}\left(\frac{\dot{a}}{a}\right)^{\alpha-2}\right].
\end{equation}
It was argued in \cite{Cai4} that for deriving the Friedmann
equations of FRW universe in general relativity, the quantity that
produces the acceleration is the active gravitational mass
$\mathcal M$ \cite{Pad2}, rather than the total mass $M$ in the
spatial region $V$. With the entropic correction term, the active
gravitational mass $\mathcal M$ will also modified as well. On one
side, from Eq. (\ref{F7}) with replacing $M$ with $\mathcal M$ we
have
\begin{equation}\label{M1}
\mathcal M =-\frac{\ddot{a}
a^2}{G}r^3\left[1-\frac{\alpha}{2}\left(\frac{r_c}{R}\right)^{\alpha-2}\right]^{-1}.
\end{equation}
On the other side, the active gravitational mass is  defined as
\cite{Cai4}
\begin{equation}\label{Int}
\mathcal M =2
\int_V{dV\left(T_{\mu\nu}-\frac{1}{2}Tg_{\mu\nu}\right)u^{\mu}u^{\nu}}.
\end{equation}
A simple calculation leads
\begin{equation}\label{M2}
\mathcal M =(\rho+3p)\frac{4\pi}{3}a^3 r^3.
\end{equation}
Equating Eqs. (\ref{M1}) and (\ref{M2}), we  find
\begin{equation}\label{addot}
\frac{\ddot{a}}{a} =-\frac{4\pi
G}{3}(\rho+3p)\left[1-\frac{\alpha}{2}\left(\frac{r_c}{R}\right)^{\alpha-2}\right].
\end{equation}
Multiplying $\dot{a}a$ on both sides of Eq. (\ref{addot}), and
using the continuity equation (\ref{Cont}) we reach
\begin{equation}\label{Fried0}
\frac{d}{dt}(\dot{a}^2) =\frac{8\pi G}{3}\frac{d}{dt}(\rho
a^2)\left[1-\frac{\alpha}{2}\left(\frac{r_c}{R}\right)^{\alpha-2}\right].
\end{equation}
Integrating of Eq. (\ref{Fried0}), we find
\begin{equation}\label{Fried1}
H^2+\frac{k}{a^2} =\frac{8\pi G}{3}\rho\left[1-\frac{\alpha}{2\rho
a^2}\left(\frac{r_c}{r}\right)^{\alpha-2} \int{\frac{d(\rho
a^2)}{a^{\alpha-2}}}\right],
\end{equation}
where $k$ is a constant of integration. Now, in order to calculate
the integral we need to find $\rho=\rho(a)$. Assuming the equation
of state parameter $w=p/\rho$ is a constant, the continuity equation
(\ref{Cont}) can be integrated immediately to give
\begin{equation}\label{rho}
\rho=\rho_0 a^{-3(1+w)},
\end{equation}
where $\rho_0$ is the present value of the energy density. Inserting
relation (\ref{rho}) in Eq. (\ref{Fried1}), after integration, we
obtain
\begin{equation}\label{Fried2}
H^2+\frac{k}{a^2} =\frac{8\pi
G}{3}\rho\left[1-\beta\left(\frac{r_c}{R}\right)^{\alpha-2} \right],
\end{equation}
where we have defined
\begin{equation}\label{beta}
\beta= \frac{\alpha}{2}\frac{ (3w+1)}{(3w+\alpha-1)}.
\end{equation}
Using Eq. (\ref{radius}), we can  rewrite Eq. (\ref{Fried2}) as
\begin{eqnarray}\label{Fried3}
&&H^2+\frac{k}{a^2}=\frac{8\pi G}{3}\rho \\ \nonumber && \times
\left[1-\beta{r_c}^{\alpha-2}
\left(H^2+\frac{k}{a^2}\right)^{\alpha/2-1} \right]
\end{eqnarray}
In the absence of the correction terms $(\alpha=0=\beta)$, one
recovers the well-known Friedmann equation in standard cosmology.
Let us note that the left hand side of Eq. (\ref{Fried3}) is always
positive thus the right hand side is also positive. This is due to
the fact that the right hand side of the usual Friedmann equation is
always positive ($\rho>0$ and $G>0$)  so $H^2+k/a^2$ should be
positive in our case to have a correct $\beta=0=\alpha$ limit. This
leads to the following condition
\begin{equation}\label{cond2}
\beta{r_c}^{\alpha-2} \left(H^2+\frac{k}{a^2}\right)^{\alpha/2-1}<1.
\end{equation}
Eq. (\ref{Fried3}) can also be written as
\begin{eqnarray}\label{Fried4}
&&\left(H^2+\frac{k}{a^2}\right)\left[1-\beta{r_c}^{\alpha-2}
\left(H^2+\frac{k}{a^2}\right)^{\alpha/2-1} \right]^{-1}\\ \nonumber
&&=\frac{8\pi G}{3}\rho.
\end{eqnarray}
Taking into account condition (\ref{cond2}) we can expand the above
equation up to the linear order of $\beta$. The result is
\begin{eqnarray}\label{Fried5}
\left(H^2+\frac{k}{a^2}\right)+\beta{r_c}^{\alpha-2}\left(H^2+\frac{k}{a^2}\right)^{\alpha/2}
=\frac{8\pi G}{3}\rho,
\end{eqnarray}
where we have neglected $O(\beta^2)$ terms and higher powers of
$\beta$. This is due to the fact that at the present time $R\gg 1$
and hence $H^2+k/a^2\ll 1 $ (see the right hand side of standard
Friedman equation where $G\sim 10^{-11}$ and $\rho\ll1$). Indeed for
the present time where the apparent horizon area (radius) becomes
large, the power-law correction terms to entropy \cite{Sau} and
hence to Friedman equation are relatively small and the usual
Friedman equation is recovered. Thus, the corrections make sense
only at the early stage of the universe where $a\rightarrow 0$. When
$a\rightarrow 0$, even the higher powers of $\beta$ should be
considered. These correction terms at the early stage of the
universe may affect on the number of e-folding during the inflation.
However this issue should be examined carefully elsewhere. With
expansion of the universe, the power-law entropy-corrected Friedmann
equation reduces to the usual Friedman equation.
%%%%%%%%%%%%%%%%%%%%%%%%%%%%%%%%%%%%%%%%%%%%%%%%%%%%%%%%%%%%%%%%%%%%%%%%%%%%%
\section{Modified Friedmann equations from the first law \label{Friedmann2}}
In this section, we adopt another approach to derive the
entropy-corrected Friedmann equation. Indeed, we are able to
derive modified Friedmann equation by applying the first law of
thermodynamics at apparent horizon of a FRW universe, with the
assumption that the associated entropy with apparent horizon has
the power-law corrected form (\ref{S}). It was already shown that
the differential form of the Friedmann equation in the FRW
universe can be written in the form of the first law of
thermodynamics on the apparent horizon \cite{Sheykhi2}. We follow
the method developed in \cite{Sheykhi3}. Throughout this section
we set $\hbar=c=k_B=1$  for simplicity. The associated temperature
with the apparent horizon can be defined as \cite{Cai5}
\begin{equation}\label{T}
T = \frac{\kappa}{2\pi}=-\frac{1}{2\pi R}\left(1-\frac{\dot R}{2H
R}\right).
\end{equation}
where $\kappa$ is the surface gravity. When $\dot {R}\leq 2H R$, the
temperature becomes negative $T\leq 0$. Physically it is not easy to
accept the negative temperature. In this case the temperature on the
apparent horizon should be defined as $T=|\kappa|/2\pi$. The work
density is obtained as \cite{Hay2}
\begin{equation}\label{Work2}
W=\frac{1}{2}(\rho-p).
\end{equation}
The work density term is regarded as the work done by the change of
the apparent horizon. We also assume the first law of thermodynamics
on the apparent horizon is satisfied and has the form
\begin{equation}\label{FL}
dE = T_h dS_h + WdV,
\end{equation}
where $S_{h}$ is the power-law corrected entropy associated with the
apparent horizon which has the form (\ref{S}). Suppose $E=\rho V$ is
the total energy content of the universe inside a $3$-sphere of
radius $R$, where $V=\frac{4\pi}{3}R^{3}$ is the volume enveloped by
3-dimensional sphere with the area of apparent horizon $A=4\pi
R^{2}$. Taking differential form of the relation $E=\frac{4\pi}{3}
\rho R^{3}$ for the total matter and energy inside the apparent
horizon, and using the continuity equation (\ref{Cont}), we get
\begin{equation}
\label{dE2}
 dE=4\pi \rho
 R^{2} dR-4\pi H R^{3}(\rho+p) dt.
\end{equation}
Taking differential form of the corrected entropy (\ref{S}), we have
\begin{equation} \label{dS}
dS_h= \frac{2\pi
R}{G}\left[1-\frac{\alpha}{2}\left(\frac{r_c}{R}\right)^{\alpha-2}\right]
dR.
\end{equation}
Inserting Eqs. (\ref{T}), (\ref{Work2}), (\ref{dE2}) and (\ref{dS})
in the first law (\ref{FL}),  we can get the differential form of
the modified Friedmann equation
\begin{equation} \label{Friedm1}
\frac{1}{4\pi
G}\frac{dR}{R^3}\left[1-\frac{\alpha}{2}\left(\frac{r_c}{R}\right)^{\alpha-2}\right]
= H (\rho+p) dt.
\end{equation}
Using the continuity equation (\ref{Cont}), we can rewrite it as
\begin{equation} \label{Friedm2}
-\frac{2}{R^3}
\left[1-\frac{\alpha}{2}\left(\frac{r_c}{R}\right)^{\alpha-2}\right]dR
= \frac{8\pi G}{3}d\rho.
\end{equation}
Integrating (\ref{Friedm2}) yields
\begin{equation} \label{Friedm3}
\frac{1}{R^2}-\frac{r_c^{\alpha-2}}{R^{\alpha}}= \frac{8\pi
G}{3}\rho+C,
\end{equation}
where $C$ is an integration constant to be determined later.
Substituting $R$ from Eq. (\ref{radius}) we obtain entropy-corrected
Friedmann equation
\begin{equation} \label{Friedm4}
H^2+\frac{k}{a^2}-r_c^{\alpha-2}\left(H^2+\frac{k}{a^2}\right)^{\alpha/2}
= \frac{8\pi G}{3}\rho+C.
\end{equation}
The constant $C$ can be determined by taking the $\alpha\rightarrow
0$ limit of the above expression. In this limit Eq. (\ref{Friedm4})
reduces to the usual Friedmann equation provided $C=-r_c^{-2}$. Thus
we reach
\begin{equation} \label{Friedm5}
H^2+\frac{k}{a^2}-r_c^{-2}\left[r_c^{\alpha}\left(H^2+\frac{k}{a^2}\right)^{\alpha/2}-1\right]
= \frac{8\pi G}{3}\rho.
\end{equation}
This is the power-law entropy corrected Friedmann equation derived
using the first law on the apparent horizon. To show the consistency
between the result of this section with Eq. (\ref{Fried5}) derived
in the previous section, let us note that Eq. (\ref{Fried5}) in the
previous section was derived for the late time where the term
$O(\beta^2)$ can be neglected and thus we do not expect to be
exactly the same as the result obtained in this section which is
valid for all epoch of the universe. However, if one absorbs, in Eq.
(\ref{Friedm4}), the constant $C$ in $\rho$, then one can rewrite
Eq. (\ref{Friedm4}) as
\begin{equation} \label{Friedm6}
H^2+\frac{k}{a^2}-r_c^{\alpha-2}\left(H^2+\frac{k}{a^2}\right)^{\alpha/2}
= \frac{8\pi G}{3}\rho,
\end{equation}
which is consistent with Eq. (\ref{Fried5}) derived using the
entropic force approach in the previous section provided one takes
$\beta=-1$, which can be translated into
\begin{equation}\label{alp}
w=\frac{2-3\alpha}{3\alpha+6}.
\end{equation}
For $\alpha>2$, the above relation leads to $w<-1/3$. Two points
should be considered here carefully. First,  relation (\ref{alp})
was derived for $\beta=-1$, thus it does not have $\alpha=0$ limit,
since in this case ($\alpha=0$), from definition (\ref{beta}) we
have $\beta=0$, which is in contradiction with condition $\beta=-1$.
Second, relation (\ref{alp}) appears when we want to show the
consistency between modified Friedman equation derived from two
methods. In the absence of correction terms $(\alpha=0=\beta)$ the
obtained Friedman equations from two different methods, namely Eqs.
(\ref{Fried5}) and (\ref{Friedm5}) exactly coincide regardless of
the value of $w$. This indicates that for usual Friedmann equation
the condition (\ref{alp}) is relaxed and hence $w$ can have any
arbitrary value in standard cosmology.

It is also notable to mention that Eq. (\ref{Friedm5}) is consistent
with the result obtained in \cite{Karami}. However, our derivation
is quite different from \cite{Karami}. Let us stress the difference
between our derivation in this section and \cite{Karami}. First of
all, the authors of \cite{Karami} have derived modified Friedmann
equations by applying the first law of thermodynamics, $TdS =-dE$,
to the apparent horizon of a FRW universe with the assumption that
the apparent horizon has corrected-entropy like (\ref{S}). It is
worthy to note  that the notation $dE$ in \cite{Karami} is quite
different from the same we used in this section. In \cite{Karami},
$-dE$ is actually just the heat flux crossing the apparent horizon
within an infinitesimal internal of time $dt$. But, here $dE$ is
change in the the matter energy inside the apparent horizon.
Besides, in \cite{Karami} the apparent horizon radius $R$ has been
assumed to be fixed. But, here,  the apparent horizon radius changes
with time. This is the reason why we have included the term $WdV$ in
the first law (\ref{FL}). Indeed, the term $4\pi R^{2}\rho d R$ in
Eq. (\ref{dE2}) contributes to the work term, while this term is
absent in $dE$ of \cite{Karami}. This is consistent with the fact
that in thermodynamics the work is done when the volume of the
system is changed.
%%%%%%%%%%%%%%%%%%%%%%%%%%%%%%%%%%%%%%%%%%%%%%%%%%%%%%%%%%%%%%%%%%%%%%%%
\section{Generalized Second law of thermodynamics\label{GSL}}
Finally, we investigate the validity of the generalized second law
of thermodynamics  for the power-law entropy corrected Friedmann
equations in  a region enclosed by the apparent horizon. Our method
here differs from that of Ref. \cite{pavon1}, in that they studied
the generalized second law along with either Clausius relation or
the equipartition law of energy, while we apply the first law of
thermodynamics (\ref{FL}). The difference between our method and
Ref. \cite{Karami} was also explained in the last paragraph of the
previous section. Substituting relation (\ref{radius}) in modified
Friedmann Eq. (\ref{Friedm5}) we find
\begin{equation} \label{GSL0}
\frac{1}{R^2}-\frac{r_c^{\alpha-2}}{R^{\alpha}}+r_c^{\alpha-2}=\frac{8\pi
G}{3}\rho
\end{equation}
Differentiating Eq. (\ref{GSL0}) with respect to the cosmic time,
after using the continuity Eq. (\ref{Cont}), we get
\begin{equation} \label{GSL1}
\dot{R}=4 \pi G H R^3 (\rho+p)\left[1-\frac{\alpha}{2}
\left(\frac{r_c}{R}\right)^{\alpha-2}\right]^{-1}.
\end{equation}
Next, we calculate $T_{h} \dot{S_{h}}$. Using Eq. (\ref{dS}) we find
\begin{equation}\label{TSh1}
T_{h} \dot{S_{h}} =\frac{1}{2\pi R}\left(1-\frac{\dot R}{2H
R}\right) \times \frac{2 \pi R}{G} \left[1-\frac{\alpha}{2}
\left(\frac{r_c}{R}\right)^{\alpha-2}\right] \dot{R}
\end{equation}
After some simplification and using Eq. (\ref{GSL1}) we obtain
\begin{equation}\label{TSh2}
T_{h} \dot{S_{h}} =4\pi H R^3 (\rho+p)\left(1-\frac{\dot R}{2H
R}\right).
\end{equation}
In the accelerating universe the dominant energy condition may
violate, $\rho+p<0$, indicating that the second law of
thermodynamics ,$\dot{S_{h}}\geq0$, does not hold. However, as we
will see below the generalized second law of thermodynamics,
$\dot{S_{h}}+\dot{S_{m}}\geq0$, is still fulfilled throughout the
history of the universe. From the Gibbs equation we have
\cite{Pavon2}
\begin{equation}\label{Gib2}
T_m dS_{m}=d(\rho V)+pdV=V d\rho+(\rho+p)dV,
\end{equation}
where $T_{m}$ and $S_{m}$ are, respectively, the temperature and the
entropy of the matter fields inside the apparent horizon. We limit
ourselves to the assumption that the thermal system bounded by the
apparent horizon remains in equilibrium so that the temperature of
the system must be uniform and the same as the temperature of its
boundary. This requires that the temperature $T_m$ of the energy
inside the apparent horizon should be in equilibrium with the
temperature $T_h$ associated with the apparent horizon, so we have
$T_m = T_h$ \cite{Pavon2}. This expression holds in the local
equilibrium hypothesis. If the temperature of the fluid differs much
from that of the horizon, there will be spontaneous heat flow
between the horizon and the fluid and the local equilibrium
hypothesis will no longer hold. Therefore from the Gibbs equation
(\ref{Gib2}) we can obtain
\begin{equation}\label{TSm2}
T_{h} \dot{S_{m}} =4\pi R^2\dot{R}(\rho+p)-4\pi R^3H(\rho+p).
\end{equation}
To check the generalized second law of thermodynamics, we have to
examine the evolution of the total entropy $S_h + S_m$. Adding
equations (\ref{TSh2}) and (\ref{TSm2}),  we get
\begin{equation}\label{GSL2}
T_{h}( \dot{S_{h}}+\dot{S_{m}})=2\pi R^{2}(\rho+p)\dot
{R}=\frac{A}{2}(\rho+p) \dot {R}.
\end{equation}
where $A=4 \pi R^2$ is the apparent horizon area. Substituting $\dot
{R}$ from Eq. (\ref{GSL1}) into (\ref{GSL2}) we find
\begin{equation}\label{GSL3}
T_{h}( \dot{S_{h}}+\dot{S_{m}})=2\pi G A H R^{3}(\rho+p)^2
\left[1-\frac{\alpha}{2}
\left(\frac{r_c}{R}\right)^{\alpha-2}\right]^{-1}.
\end{equation}
As we argued after Eq. (\ref{F5}) the expression in the bracket of
Eq. (\ref{GSL3}) is always positive i.e.,
\begin{equation}\label{cond}
\left[1-\frac{\alpha}{2}
\left(\frac{r_c}{R}\right)^{\alpha-2}\right] >0.
\end{equation}
Thus the right hand side of Eq. (\ref{GSL3}) cannot be negative
throughout the history of the universe, which means that $
\dot{S_{h}}+\dot{S_{m}}\geq0$ always holds. This implies that for a
universe with power-law entropy corrected relation the generalized
second law of thermodynamics is fulfilled in a region enclosed by
the apparent horizon. Note that if we identify the crossover scale
$r_c$ with the present value of the apparent horizon, i.e., $r_c=R$,
then the condition (\ref{cond}) reduces to $\alpha<2$, which is
consistent with the result obtained in \cite{pavon1,Karami}.

%%%%%%%%%%%%%%%%%%%%%%%%%%%%%%%%%%%%%%%%%%%%%
\section{Conclusions and discussions\label{Con}}
It was argued that a possible source of black hole entropy could
be the entanglement of quantum fields in and out the horizon
\cite{Sau}. The entanglement entropy of the ground state of field
obeys the well-known area law. However, the power-law correction
to the area law appears when the wave-function of the quantum
field is chosen to be a superposition of ground state and exited
state \cite{Sau}. Indeed, the excited states contribute to the
correction, and more excitations produce more deviation from the
area law \cite{sau1,sau2}. Therefore, the correction terms are
more significant for higher excitations.

Motivated by the power-law corrected entropy and adopting the
viewpoint that gravity emerges as an entropic force, we derived
modified Newton's law of gravitation as well as power-law correction
to Friedmann equations. We found that the correction term for
Friedmann equation falls off rapidly with apparent horizon radius
and can be comparable to the first term only when the scale factor
$a$ is very small. Thus the corrections make sense only at early
stage of the universe. When the universe becomes large, the
power-law entropy-corrected Friedmann equation reduces to the
standard Friedman equation. This can be understood easily.  At late
time where $a$ is large, i.e., at low energies, it is difficult to
excite the modes and hence, the ground state modes contribute to
most of the entanglement entropy. However, at the early stage, i.e.,
at high energies, a large number of field modes can be excited and
contribute significantly to the correction causing  deviation from
the area law and hence deviation from the standard  Friedmann
equation.

We also derived modified Friedmann equation from different approach.
Starting from the first law of thermodynamics at apparent horizon of
a FRW universe, and assuming that the associated entropy with
apparent horizon has power-law corrected form (\ref{S}), we obtained
modified Friedmann equation. We find out that the derived modified
equations from these two different approaches (entropic force
approach and first law approach) can be consistent provided the
equation of state parameter satisfies in condition (\ref{alp}).
However, in the absence of the correction terms $(\alpha=0=\beta)$
the obtained Friedman equations from two different methods, namely
Eqs. (\ref{Fried5}) and (\ref{Friedm5}) exactly coincide regradless
of the value of $w$. This indicates that for usual Friedmann
equation the condition (\ref{alp}) is relaxed and hence $w$ can have
any arbitrary value in standard cosmology.

Finally, we investigated the validity of the generalized second law
of thermodynamics for the FRW universe with any spatial curvature.
We have shown that, when thermal system bounded by the apparent
horizon remains in equilibrium with its boundary such that $T_m =
T_h$, the generalized second law of thermodynamics is fulfilled in a
region enclosed by the apparent horizon. The results obtained here
for power-law corrected entropy area relation further supports the
thermodynamical interpretation of gravity and provides more
confidence on the profound connection between gravity and
thermodynamics.
%%%%%%%%%%%%%%%%%%%%%%%%%%%%%%%%%%%%%%%%%%%%%%%%%%%%%%%%%%%%%%%%%%%%%%%
\acknowledgments{We thank the anonymous referee for constructive and
valuable comments. This work has been supported by Research
Institute for Astronomy and Astrophysics of Maragha.}
%%%%%%%%%%%%%%%%%%%%%%%%%%%%%%%%%%%%%%%%%%%%%%%%%%%%%%%%%%%%%%%%%%%%%%%%%%%

\end{document}